\documentclass[twocolumn,preprintnumbers,nofootinbib,prd,superscriptaddress]{revtex4}
\pdfoutput=1

\usepackage[utf8]{inputenc}
\usepackage{graphicx}
\usepackage{amssymb}
\usepackage{amsxtra}
\usepackage{amsmath}
\usepackage{booktabs,multirow,tabularx}
\usepackage{slashed}
\usepackage{float}
\usepackage{placeins}
\usepackage{rotating}
\usepackage{lscape}
\usepackage{color}
\usepackage{hyperref}

\usepackage[Q=yes,pverb-linebreak=no]{examplep}

\DeclareGraphicsExtensions{.pdf}
\graphicspath{{./}}

\newcommand{\vev}[1]{\langle {#1} \rangle}
\newcommand{\lsim}{\lesssim}

\newcommand{\ord}[1]{\mathcal{O}{(#1)}}
\newcommand{\eq}[1]{Eq.~(\ref{#1})}
\newcommand{\fig}[1]{Fig.~(\ref{#1})}

\newcommand{\gev}{\,\textrm{GeV}}

\def\beq{\begin{equation}}
\def\bea{\begin{eqnarray}}
\def\eeq{\end{equation}}
\def\eea{\end{eqnarray}}
\def\beqnl{\begin{align}}
\def\endal{\end{align}}

\newcommand{\suD}{SU(3)_D}
\newcommand{\ztwo}{{\mathbb Z_2}}
\DeclareFontFamily{U}{cbgreek}{}
\DeclareFontShape{U}{cbgreek}{m}{n}{
        <-6>    grmn0500
        <6-7>   grmn0600
        <7-8>   grmn0700
        <8-9>   grmn0800
        <9-10>  grmn0900
        <10-12> grmn1000
        <12-17> grmn1200
        <17->   grmn1728
      }{}
\DeclareFontShape{U}{cbgreek}{bx}{n}{
        <-6>    grxn0500
        <6-7>   grxn0600
        <7-8>   grxn0700
        <8-9>   grxn0800
        <9-10>  grxn0900
        <10-12> grxn1000
        <12-17> grxn1200
        <17->   grxn1728
      }{}

\makeatletter
\newcommand{\normalorbold}{%
  \ifnum\pdf@strcmp{\math@version}{bold}=\z@ bx\else m\fi
}
\makeatother

\begin{document}
\title{\boldmath Unified Scenario for Composite Right-Handed Neutrinos and Dark Matter}

\author{Hooman Davoudiasl\footnote{email: hooman@bnl.gov}}

\author{Pier Paolo Giardino\footnote{email: pgiardino@bnl.gov}}

\affiliation{Department of Physics, Brookhaven National Laboratory, Upton, NY 11973, USA}

\author{Ethan T. Neil\footnote{email: ethan.neil@colorado.edu}}

\affiliation{Department of Physics, University of Colorado, Boulder, Colorado 80309, USA}
\affiliation{RIKEN-BNL Research Center, Brookhaven National Laboratory, Upton, New York 11973, USA}

\author{Enrico Rinaldi\footnote{email: erinaldi@bnl.gov }}

\affiliation{RIKEN-BNL Research Center, Brookhaven National Laboratory, Upton, New York 11973, USA}

\begin{abstract}
We entertain the possibility that neutrino masses and dark matter (DM) originate from a common composite dark sector.
A minimal effective theory can be constructed based on a dark $\suD$ interaction with three flavors of massless dark quarks;
electroweak symmetry breaking gives masses to the dark quarks.
By assigning a $\ztwo$ charge to one flavor, a stable ``dark kaon'' can provide a good thermal relic DM candidate.
We find that ``dark neutrons'' may be identified as right handed Dirac neutrinos.
Some level of ``neutron-anti-neutron'' oscillation in the dark sector can then result in non-zero Majorana masses for light Standard Model neutrinos.
A simple ultraviolet completion is presented, involving additional heavy $\suD$-charged particles with electroweak and lepton Yukawa couplings.
At our benchmark point, there are ``dark pions'' that are much lighter than the Higgs and we expect spectacular collider signals arising from the UV framework.
This includes the decay of the Higgs boson to $\tau \tau \ell \ell^{\prime}$, where $\ell$($\ell'$) can be any lepton, with displaced vertices.
We discuss the observational signatures of this UV framework in dark matter searches and primordial gravitational wave experiments;
the latter signature is potentially correlated with the $H \to \tau \tau \ell \ell^{\prime}$ decay.
\end{abstract}

\maketitle

\section{Introduction\label{sec:intro}}

The Standard Model (SM) of particle physics encodes our most precise description of microscopic phenomena.
The discovery of the Higgs scalar in 2012 provided its last missing piece.
For decades, the SM has withstood a variety of precision tests and so far there is no definitive experimental deviation from its predictions.  While this situation could possibly change over the next few years - with more data becoming available at the LHC and elsewhere - there are already robust reasons to conclude that the SM offers only a partial description of Nature.

Among the most convincing reasons for this conclusion are neutrino flavor oscillations, requiring non-zero neutrino masses $m_\nu \neq 0$, and the overwhelming body of evidence for an unknown substance, dark matter (DM), which makes up about a
quarter of the energy density in the Universe~\cite{Olive:2016xmw}.
Given that there are no feasible ways within the SM to account for these two observations, one is inexorably led to invoke new physics in order to explain them.

To explain $m_\nu\neq 0$, obviously one needs to introduce new states that have some coupling with the SM.
Very often, a successful DM model also requires some level of interaction between the new particles and the SM.
In both cases, the assumed interactions are expected to be fairly weak, given the small inferred values of $m_\nu$ and the elusive nature of DM.
However, typically these two phenomena are assumed to have different origins.

In this work, we consider the possibility that both DM and neutrino masses are manifestations of the same underlying dynamics coming from a `dark sector' that only has suppressed couplings with the SM.
The new requisite states are composite and emerge after the dark sector confines, with the confining dynamics associated with a $\suD$ gauge group and three flavors of Dirac fermions in the fundamental representation, {\it i.e.} the dark quarks.\footnote{For a recent paper where neutrino masses are related to a new invisible sector similar to the SM see Ref.~\cite{Berryman:2017twh}. A warped seesaw mechanism, holographically dual to a composite model, has been studied in Ref.~\cite{Agashe:2015izu}.}

Given the above ingredients, the role of `right-handed neutrinos' that mediate a mechanism for neutrino mass generation is played by the low-lying composite dark baryons, close analogues of ordinary QCD nucleons.
Since the dark baryons couple to SM states, they cannot be stable enough to be DM.
However, assuming that one of the dark flavors is odd under a $\ztwo$, the model naturally has a stable DM candidate.\footnote{A recent review focusing on the nature of composite dark matter and its signatures can be found in Ref.~\cite{Kribs:2016cew}.}
By analogy with the SM, we will refer to the right-handed neutrino states as ``dark neutrons'', since they have no charge under the assumed interactions and symmetries.
In our model, the light SM neutrinos with $m_\nu\lsim 0.1 $~eV are Majorana states.
This requires that the dark neutrons, which are Dirac fermions, have a small Majorana mass, induced by the dark sector analogue of ``neutron-anti-neutron oscillation'' (NANO) operators.

In the setup described so far, the requisite couplings to the SM neutrinos and NANO effects descend from higher-dimension operators involving dark quarks.
Upon confinement, these operators furnish the dimension-4 interactions that involve SM states, dark baryons, and other dark hadrons.
The new particles can be assumed to have masses well above the weak scale.
However, as we will discuss later, they could potentially be within the reach of the LHC or future envisioned high energy colliders.
Next, we will introduce the effective theory for the $\suD$ theory interactions with the SM.
Then we will also introduce a simple underlying model that will provide the ultraviolet (UV) completion of the effective theory described by the higher-dimension operators.

\section{Effective Theory\label{sec:ET}}

Let $\psi_i$ be three Dirac fermions in the fundamental representation of an $SU(n_c)_D$ `dark' gauge interaction with no SM charges.
We will take these fermions to be exactly massless in the high-energy theory, with the corresponding chiral symmetries broken by effective Higgs interactions after electroweak symmetry breaking.
The addition of explicit vector-like mass terms can lead to interesting variations, but does not change the basic structure of our model.

Since we will later use the composite baryons of this theory as {\it fermionic} partners of left-handed SM neutrinos $\nu_L$, the number of colors $n_c$ must be an odd integer: for simplicity we will focus on $n_c=3$.
Our high-energy theory so far is therefore a dark analogue of QCD, with massless quarks.
In addition to the symmetries of QCD, we introduce an additional $\ztwo$ symmetry which will stabilize one of the composite dark hadrons on cosmological time scales.
Under this symmetry, only $\psi_3$ is odd: $\ztwo (\psi_3) = -1 = - \ztwo (\psi_{1,2})$, and the symmetry is not anomalous, because it acts on both chiralities of $\psi_3$.

We will not write down the Lagrangian ${\cal L}_{\rm DQCD}$ for the $\suD$ theory, as it is well known and identical
to the SM QCD with three flavors.
We will assume that this theory is augmented by three types of higher-dimensional operators ${\cal L}={\cal L}_{\rm DQCD}+{\cal L} _{\rm eff}$, deriving from different high-energy effective scales $\Lambda_X$ and schematically given by
\beq
{\cal L} _{\rm eff} = \frac{{\tilde H^*} \bar L_f \, [\psi_i^3]}
{\Lambda_f^3} + \frac{[\psi_i^6]}{\Lambda_N^5} + \frac{\bar \psi_i \psi_i H^\dagger H}{\Lambda_H} + {\small \rm H.C.}\,,
\label{ET}
\eeq
where ${\tilde H^*} = \epsilon_{a b} H^{a *}$, with $H$ the SM Higgs doublet, and $L_f$ is a lepton doublet of the SM family $f=1,2,3$.
In \eq{ET} $i=1,2,3$ and $[\psi_i^n]$ represents any $\suD$ singlet and Lorentz invariant combinations of $n$ $\psi_i$ quarks that are $\ztwo$ even.

Electroweak symmetry breaking will also break the chiral symmetries related to massless $\psi_i$, yielding masses for the dark fermions from the third term in \eq{ET}
\beq
m_i \sim \frac{\vev{H}^2}{\Lambda_H},
\label{psimass}
\eeq
where $\vev{H}= v_H/\sqrt{2}$ with $v_H \approx 246$~GeV.
In the following we assume the mass ordering $m_1 < m_2 < m_3$, with masses of roughly the same order and much smaller than the scale $\mu_D$ of dark confinement: $m_i \ll \mu_D$.

In what follows, we will discuss the qualitative features of our scenario.
We will also provide rough estimates for the expected sizes of various effects that arise as a result of $\suD$ confinement.
These estimates are only meant to be treated as order-of-magnitude guides.
In fact, more precise quantitative results would require non-perturbative lattice field theory computations, for any specific choice of parameters.
We begin with the generation of neutrino mass from mixing with ``dark neutron'' operators, and then discuss a plausible scenario for composite dark matter stabilized by the $\ztwo$ symmetry of $\psi_3$.

\subsection{Low-energy spectrum \label{sec:spectrum}}

Below $\mu_D$, the operators $[\psi_i^n]$ will transmute into various dark hadronic operators.
To fully understand the low-energy Lagrangian in terms of these dark hadrons, we begin by mapping out the expected spectrum of low-lying bound states, which will be qualitatively very similar to QCD since we have the same gauge group [$SU(3)$] and number of light fermion species (three).

With our assumption $m_i \ll \mu_D$, we expect chiral perturbation theory to be applicable to this model.
The chiral Lagrangian predicts an octet of (pseudo-)Nambu-Goldstone bosons (pNGB): the three states
\beq
P \sim \bar{\psi}_1 \psi_2,\ \kappa \sim \bar{\psi}_1 \psi_3,\ \kappa' \sim \bar{\psi}_2 \psi_3,
\eeq
their antiparticles, and two linear combinations $P'$ and $P''$ of the flavor-diagonal bilinear $\bar{\psi}_i \psi_i$, analogues of the $\pi^0$ and $\eta$ in QCD.
We will denote these pNGB states collectively with the symbol $\Pi$.
All of these states are $\ztwo$-even except for the ``dark kaons'' $\kappa$ and $\kappa'$, containing only one $\psi_3$.
The dark kaons are the lightest $\ztwo$-odd hadrons and $\kappa$ will provide a dark matter candidate in this model.

At heavier masses in the confined spectrum, we expect a number of baryon-like bound states.
The most important ones for our purposes are the spin-$1/2$, $\ztwo$-even states - the ``dark neutrons'' - which will allow neutrino mass generation after electroweak symmetry breaking:
\beq
N_1 \sim \psi_1^2 \psi_2,\ N_2 \sim \psi_1 \psi_2^2,\ N_3 \sim \psi_1 \psi_3^2,\ N_4 \sim \psi_2 \psi_3^2.
\label{neutrons}
\eeq
With the given dark fermions mass ordering, we have Dirac masses for the dark neutrons  satisfying $m_{N_1} < m_{N_2} < m_{N_3} < m_{N_4}$, and of the order of the confinement scale.
As in QCD, there will be small contributions from the explicit fermion masses $m_i$, but overall we take $M_N \sim \mu_D$ (which amounts to the identification: $\Lambda_{\rm QCD}\sim 1$~GeV in QCD.)
Most of the other hadronic resonances in the dark sector will also have masses of order the confinement scale $\mu_D$, except for the pNGB states $\Pi$.
We can estimate the mass of the latters using the Gell-Mann-Oakes-Renner relation~\cite{GellMann:1968rz}, predicting
\beq
M_{\Pi}^2 = 2b \mu_D \hat{m}\, ,
\eeq
where $\hat{m}$ is the average dark fermion mass, and $b$ is a dimensionless low-energy constant.
In QCD, the average of the down and strange quark masses is $(m_d + m_s)/2 \approx 50$ MeV, while the neutral kaon mass is approximately 500 MeV, so taking $\Lambda_{\rm QCD} \sim 1$ GeV, we identify $b \sim 2.5$, or
\beq
M_{\Pi} \sim  \sqrt{5 \mu_D \hat{m}} \sim \vev{H} \sqrt{\frac{5\mu_D}{\Lambda_H}} \, ,
\eeq
using \eq{psimass}.
This provides the basis of our benchmark point, where we take $\mu_D \sim 1$~TeV and $\Lambda_H \sim 1000$~TeV, yielding $M_{\Pi} \sim 10$ GeV.
This choice is anticipating the value of $M_{\kappa} \sim M_{\Pi}$ which will be consistent with the required separation of scales to achieve the correct dark matter relic density, as we will show in Sec.~\ref{sec:relic-density}.
If a single UV completion gives rise to all of the effective scales $\Lambda_X$, then we are assuming some relative suppression in the generation of the mass operator $\bar{\psi}_i\psi_iH^\dagger H$.

Among the other possible three-fermion bound states, $\psi_1^3$ and $\psi_2^3$, although $\ztwo$-even, must be spin-$3/2$ due to the total antisymmetry of their wavefunctions under identical fermion exchange (similar to the $\Delta^{++}$ resonance in QCD.)
These spin-$3/2$ fermions cannot mix directly with neutrinos to give rise to neutrino masses, and we will not consider them further.

Finally, we expect a number of $\ztwo$-odd baryon-like bound states:
\beq
X \sim \psi_1^2 \psi_3,\ X' \sim \psi_2^2 \psi_3,\ Y_{3/2} \sim \psi_3^3,
\eeq
where the third state has spin-$3/2$.
Since we have only a $\ztwo$ symmetry, with the U$(1)$ baryon number of the individual $\psi_i$ broken in our effective Lagrangian \eq{ET}, these heavy baryons are all expected to be unstable and decay into mesons, including at least one of the $(\kappa, \kappa')$, in order to preserve $\ztwo$.
We note in passing that if large vector-like masses were added to the theory so that $m_3 \ll m_1, m_2$, then it would be possible for $Y_{3/2}$ to be the lightest $\ztwo$-odd hadron instead, possibly leading to a model of composite spin-$3/2$ dark matter.
We will not pursue this scenario further in this work.

Note that strong dynamics will not break the $U(1)$ flavor symmetry associated with
each $\psi_i$~\cite{Vafa:1983tf}.
Hence, $\vev{\bar \psi_1 \psi_3} = \vev{\bar \psi_3 \psi_1} = 0$, and similarly for $\vev{\bar \psi_3 \psi_2}$, {\it i.e.} it also respects the $\ztwo$ symmetry.
This is similar to how QCD only has flavor-diagonal condensates and does not lead to the breaking of electromagnetic $U(1)_{\rm EM}$.

In addition to the aforementioned bound states, the confining nature of the dark sector can accommodate glueballs and other exotic states.
In the presence of dark quarks, glueballs are not stable and are heavy according to lattice QCD calculations~\cite{Gregory:2012hu,Chen:2005mg} with their mass at the confinement scale.\footnote{For theories of dark matter without fermions, where the lightest scalar glueball is the dark matter candidate see for example Refs.~\cite{Soni:2016gzf,Boddy:2014yra}}

\subsection{Neutrino mass generation \label{sec:neutrino-mass}}

Passing from the effective operator description of \eq{ET} to a low-energy description in terms of the ``dark neutrons'' $N_\alpha$ below the confinement scale $\mu_D$, we will have
\beq
[\psi_i^3] \to \mu_D^3 N_\alpha + \ldots\,,
\label{psi3}
\eeq
where $\alpha=1,...,4$ labels the hadrons in \eq{neutrons}.
The ellipsis in \eq{psi3} correspond to other composite operators that descend from $[\psi_i^3]$.

Similarly,
\beq
[\psi_i^6] \to \mu_D^6 N_\alpha N_\beta + \ldots\,.
\label{psi6}
\eeq

We see that Eqs.~(\ref{psi3}) and (\ref{psi6}) imply that after confinement, the Lagrangian (\ref{ET}) will yield, among others, the following terms
\beq
M_\alpha \bar N_\alpha N_\alpha + Y^{f \alpha} \, {\tilde H^*} \bar L_f \, N_\alpha +
\mu^{\alpha \beta}_N \bar N_\alpha^c N_\beta + {\small \rm H.C.}\,,
\label{NLag}
\eeq
in the low-energy description in terms of dark hadrons.
In \eq{NLag}, $M_\alpha$ is the Dirac mass of $N_\alpha$ baryons, $Y^{f \alpha}$ is a Yukawa coupling matrix, and $\mu^{\alpha\beta}_N$ is a Majorana mass matrix for $N_\alpha$ that leads to NANO.
The parameters in \eq{NLag} are estimated to be
\beq
M_\alpha \sim M_N \sim \mu_D \, ; \quad
Y^{f \alpha} \sim \frac{\mu_D^3}{\Lambda_f^3} \, ; \quad
\mu^{\alpha\beta}_N \sim \frac{\mu_D^6}{\Lambda_N^5}\,.
\label{MYmN}
\eeq

\begin{figure}[ht]
\includegraphics[width=0.48\textwidth]{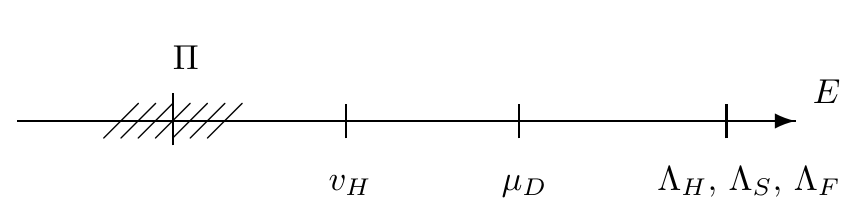}
\caption{The figure shows the separation of energy scales, from the spectrum of pNGBs $\Pi$, to the electroweak and confinement scales, all the way up to the UV region.\label{fig:scales}}
\end{figure}

Our goal is to reproduce the SM neutrino masses and we need to consider what type of parameters in our effective description would yield viable values of $m_\nu$.
Given the interactions in \eq{NLag}, the light SM neutrinos will have masses
\beq
m_\nu \sim \frac{y^2 v_H^2 \mu_N}{M_N^2}\,,
\label{mnu}
\eeq
where $y$ and $\mu_N$ are typical eigenvalues of $Y^{f \alpha}$ and $\mu_N^{\alpha\beta}$, respectively.
Substituting in the parametric dependence given in \eq{MYmN}, we have
\beq\label{nu-mass}
m_\nu \sim \frac{\mu_D^{10} v_H^2}{\Lambda_f^6 \Lambda_N^5}.
\eeq
For example, extending our benchmark point which had $\mu_D \sim 1$ TeV, we see that SM neutrino masses $m_\nu \sim 0.1$~eV are obtained if $\Lambda_f$ and $\Lambda_N$ are generated at a scale of $\ord{10}$~TeV.
This corresponds to a dark neutron Majorana mass of order $\mu_N \sim 10$ MeV and a Yukawa coupling $y \sim 10^{-3}$.
Note that these values of $\Lambda_{N,f}$ relative to $\mu_D$ are consistent with a valid effective field theory interpretation for \eq{ET}.
A sketch of the separation of energies, from the pNGB mesons to the UV scales, is presented in Fig.~\ref{fig:scales}.

\subsection{Dark matter relic density \label{sec:relic-density}}

We now move on to consider our second goal of reproducing the current dark matter abundance.
We consider a thermal relic scenario for $\kappa$.
The annihilation process $\kappa^\dagger \kappa \leftrightarrow \Pi^\dagger \Pi$ will keep $\kappa$ in equilibrium with the other pNGBs.
The thermally averaged cross section for this process will then determine the $\kappa$ relic density.
This is true as long as the decay width for lighter $\ztwo$-even $\Pi$ into SM final states is greater than the Hubble scale $H_\star$ at which the $\kappa$ annihilation freezes out\footnote{In Sec.~\ref{sec:dark-pion} we show that this condition is satisfied in a concrete UV model.}.
$H_\star$ is defined as $H_\star \sim g_{\star}^{1/2} T^2_\star/M_{\rm Planck}$, with the Planck mass $M_{\rm Planck} \sim 10^{19}$~GeV, and $ g_{\star} \sim 50$ at the freeze-out temperature $T_{\star} = M_{\kappa}/ x_{f.o.} \sim 1 \gev$, where we used $x_{f.o.} \sim 20$.

Although the decay of the $\Pi$ states allows the number density of $\kappa$ to be efficiently transferred into SM final states, the inverse decay process is not sufficient to keep the dark sector in thermal equilibrium with the Standard Model until $\kappa$ freeze-out occurs, due to Boltzmann suppression.
However, thermal contact between the dark sector and the SM also occurs through the effective operator $\bar{\psi}_i \psi_i H^\dagger H / \Lambda_H$ from \eq{ET}, responsible for giving dark fermions mass.
This dimension-5 operator will induce an interaction
\beq \label{eq:pipi}
\mathcal{L}  \supset \xi  \Pi^\dagger \Pi H^\dagger H\,,
\eeq
where $H^\dagger H$ couples as a scalar current to the pNGBs.
In the framework of chiral perturbation theory, it will enter in the same way as the mass term $m_i$.
This means that at leading order the same low-energy constant $b \sim 2.5$ identified above in the discussion of $M_\Pi$ will appear in the coupling $\xi$; we estimate
\beq
\xi \sim \frac{2 b \mu_D}{\Lambda_H}.
\eeq

The interaction in \eq{eq:pipi} allows the process between pNGBs and the charm quark, $\Pi \ c \leftrightarrow \Pi \ c$ and one can show\footnote{The scattering cross-section is $\sigma_{\Pi c} \sim 3/(16\pi) \xi^2 y_c^2 v_H^2/M_H^4$, where 3 counts the QCD colors, $y_c$ is the charm Yukawa coupling and $M_H$ the Higgs mass. Thus the rate at the freeze-out temperature is $\sim \sigma_{\Pi c} T_{\star}^{3} e^{-M_{\Pi}/T_{\star}}$ which, for our choice of parameters, is $\sim H_\star$.} that this will mantain thermal contact between the two sectors all the way down to the charm threshold $\sim 1 \gev \sim T_{\star}$.

The above discussion implies that we can use the results described in Ref.~\cite{Buckley:2012ky} for a different model of meson-like composite dark matter based on a dark sector with $SU(2)$ gauge group.
Our model has several important differences which can change the precise numerical results of the reference, but we expect the qualitative result to be unmodified, namely that for meson-like composite dark matter, the freeze-out of pNGB annihilation yields the correct dark matter relic density for the parameter choice
\beq
M_\Pi \sim M_{\kappa} \lesssim F_\Pi \sim \frac{\mu_D}{4\pi}.
\eeq
This is satisfied well by our benchmark point, with $M_\Pi \sim 10$ GeV and $F_\Pi \sim 100$ GeV.
A detailed calculation in this model would yield a more precise relationship between $M_\Pi$ and $\mu_D$ in order for the correct relic density to be achieved, but it goes beyond the scope of this paper, where only order-of-magnitude estimates are provided.

\section{UV Framework \label{sec:UV}}

Here, we present a simple UV model, intended as a basic framework that could yield the higher-dimension terms of the effective theory in \eq{ET}.
Although other UV completions are certainly possible, our choice naturally accommodates our benchmark effective theory parameters, and leads to several interesting experimental signatures.

Let $F$ and $F'$ be vector-like fermions with the quantum numbers $(3, 2, -1/2, +)$ and $(3, 2, -1/2, -)$ under $\suD \times SU(2)_L \times U(1)_Y\times \ztwo$, respectively.
We also introduce three scalars $S_a$, with $a=e,\mu,\tau$, that are triplets under $\suD$, with quantum numbers $(3,1,0, +)$.
The UV Lagrangian is taken to be
\bea \label{LagUV}
{\cal L}_{\rm UV} &=& \lambda_1 {\tilde H^*} \bar F  \psi_1 + \lambda_2 {\tilde H^*} \bar{F} \psi_2 + \lambda_3 {\tilde H^*} \bar F'  \psi_3 \\ \nonumber
&+& \lambda'_a S_a \bar F L_a
+ \sum_{i,j=1}^{3} g^{ij}_{a} S_a \psi_i \psi_j\Big|_{\ztwo=+} \\ \nonumber
&+& \mu_S S_e S_\mu S_\tau + {\small \rm H.C.} \,,
\eea
where $\lambda_i$, $\lambda'_a$ and $g^{ij}_{a}$ are coupling constants and $\mu_S$ is a mass that characterizes the $S_a$ triple coupling;
contractions of color charges for $S_a$ and $\psi_i$ are assumed.
All terms respect $\ztwo$ symmetry.

We impose a generational structure on the coupling of $S_a$ to leptons, which is implied by our choice of index label.
This is required for the model to be viable, since anarchic couplings between $S_a$ and the three lepton families generally leads to unacceptably large contributions to flavor-changing processes such as $\mu^- \rightarrow e^- \gamma$.
The antisymmetry of the $S^3$ vertex under $\suD$ then requires the given structure including all three generations at once.

\begin{figure}[ht]
\includegraphics[width=0.48\textwidth]{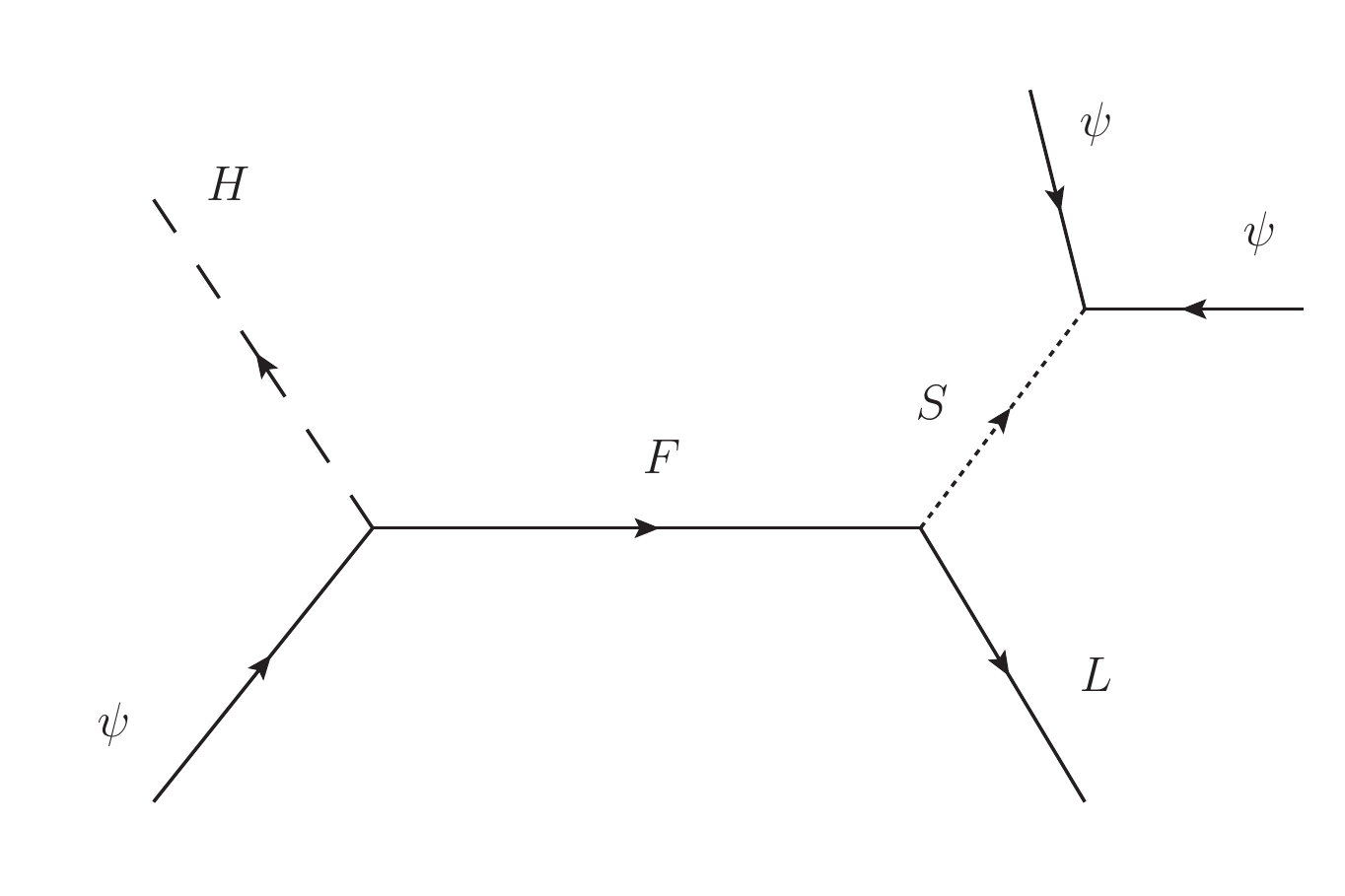}
\caption{Schematic Feynman diagram showing how the neutrino mass operator $\tilde{H}^* \bar L [\psi_i^3] / \Lambda_f^3$ is generated from our UV completion (``reindeer'' diagram). \label{fig:reindeer}}
\end{figure}

\begin{figure}[ht]
\includegraphics[width=0.48\textwidth]{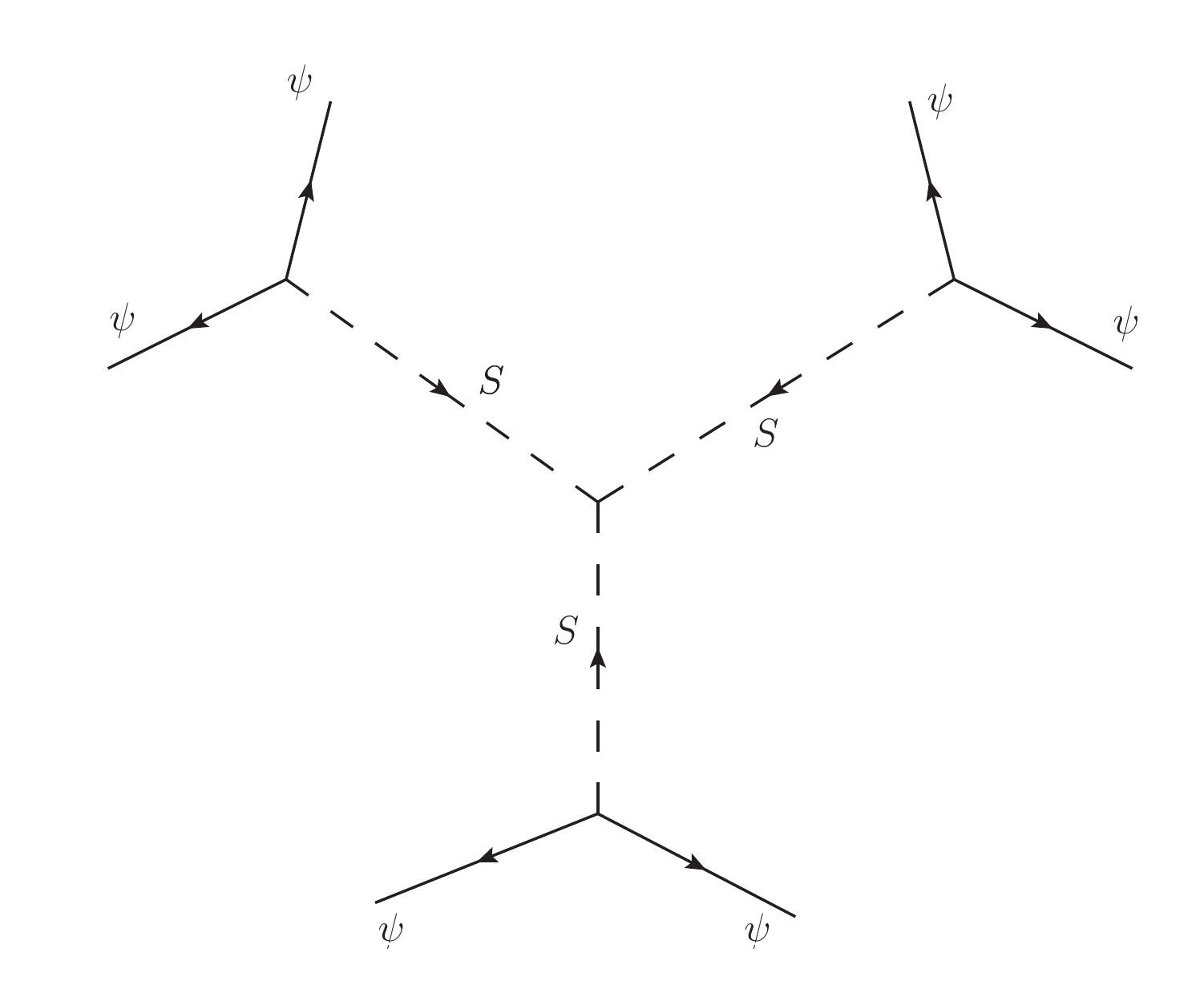}
\caption{Schematic Feynman diagram showing how the NANO operator $[\psi_i]^6 / \Lambda_N^5$ is generated from our UV completion (``snowflake'' diagram). \label{fig:snowflake}}
\end{figure}

The effective interactions in Lagrangian (\ref{ET}) can then be derived from the renormalizable terms in ${\cal L}_{\rm UV}$.
One finds
\beq \label{Lambdas}
\Lambda_f^{-3} \sim \frac{\lambda_i\, \lambda'_a\, g_{j a}}{M_F \,M_S^2} \, ; \quad
\Lambda_N^{-5} \sim \frac{g_{i a}^3  \, \mu_S}{M_S^6} \, ; \quad
\Lambda_H^{-1} \sim \frac{\lambda_i^2}{M_F} \, ,
\eeq
where $M_F$ and $M_S$ are the masses of the $F$ and $S_a$ fields, respectively (we have assumed a universal mass for $S_a$ scalars, for simplicity).
Schematically, we see that the first and second terms in the effective Lagrangian (\ref{ET}) are generated by the diagrams in \fig{fig:reindeer} (the ``reindeer'' diagram) and \fig{fig:snowflake} (``snowflake'' diagram), respectively.
The third operator which gives mass to the $\psi_i$ is generated by exchange of a single $F$ or $F'$ fermion.

We note in passing that another ``snowflake'' diagram similar to \fig{fig:snowflake} will produce the effective theory operator $\bar{L}_e \bar{L}_\mu \bar{L}_\tau FFF / \Lambda_\Psi^5$, with particular flavor structure imposed by the form of the triple-$S$ vertex.  This operator has an extremely unusual structure, and can mediate a three-generation lepton decay of the triply-charged $F^+ F^+ F^+$ baryon into $e^+ \mu^+ \tau^+$.  In our benchmark model, the rate for this process is highly suppressed compared to a decay mediated instead by the operator $\tilde{H^*} \bar{F} \psi_i$ into a doubly-charged baryon and a $W$ boson, but it could provide an intriguing signature in variations of this model.

For our benchmark point, taking $M_F \sim M_S \sim 1$ TeV leads to the identification $\lambda_i^2 \sim 10^{-3}$.
To generate the remaining scales at the benchmark, we choose $\lambda'_i \sim 0.1$, $g_i \sim 0.3$ and $\mu_S \sim 1$ GeV;
some adjustment would be possible without disturbing the effective theory benchmark point, since we have three parameters and only two scales to generate.
We summarize our benchmark set of UV parameters and the resulting set of low-energy effective couplings in Table~\ref{tab:benchmark}.

\begin{table}
  \begin{tabular}{|r@{$\sim$}l||r@{$\sim$}l|}
    \hline
    $\Lambda_H$ & 1000 TeV                  & $\lambda_i^2$ & 0.001 \\
    $\Lambda_f$, $\Lambda_N$& 10 TeV & $\lambda'_i$ & 0.1 \\
    $\mu_D$ & 1 TeV                              & $g_i$ & 0.3 \\
    $M_F$, $M_S$ & 1 TeV                       & $\mu_S$ & 1 GeV \\
    $M_{\kappa}$ & 10 GeV                        & \multicolumn{2}{c|}{}\\
    \hline
  \end{tabular}
  \caption{Summary of the scales and UV couplings at our benchmark point. The mass of our dark matter candidate $M_{\kappa}$ is also included.\label{tab:benchmark}}
\end{table}

\subsection{Dark pion decay \label{sec:dark-pion}}

We now turn to the decay of dark hadrons not stabilized by $\ztwo$ symmetry.
Due to the violation of baryon number in our effective theory, all of the baryon-like dark hadrons are expected to decay promptly into pNGBs.

Before we discuss the decay modes induced by the UV completion, we note that in the low-energy effective theory, the presence of an interaction $\Pi\, \bar{N}_\alpha \gamma_5 N_\beta$ (as in QCD) coupled with the operator $Y^{f\alpha} \bar{H}^* \bar{L}_f N_\alpha$ leads to the tree-level decay of $\Pi$ into various final states, dominantly $\Pi \rightarrow \bar{\nu} \nu \bar{b} b$\footnote{Decay into the two-body final state $\Pi \rightarrow \bar{\nu} \nu$ is negligible due to helicity suppression; the spin-zero initial state requires emission of the neutrinos with zero total angular momentum, requiring a mass flip which suppresses the decay by $(m_\nu / M_\Pi)^2$.}.
For our benchmark point we estimate this decay rate to be far too small to couple the $\Pi$ to the SM heat bath as required by our relic density model, so the dominant $\Pi$ decay process must arise from our UV completion.

\begin{figure}[ht]
\includegraphics[width=0.45\textwidth]{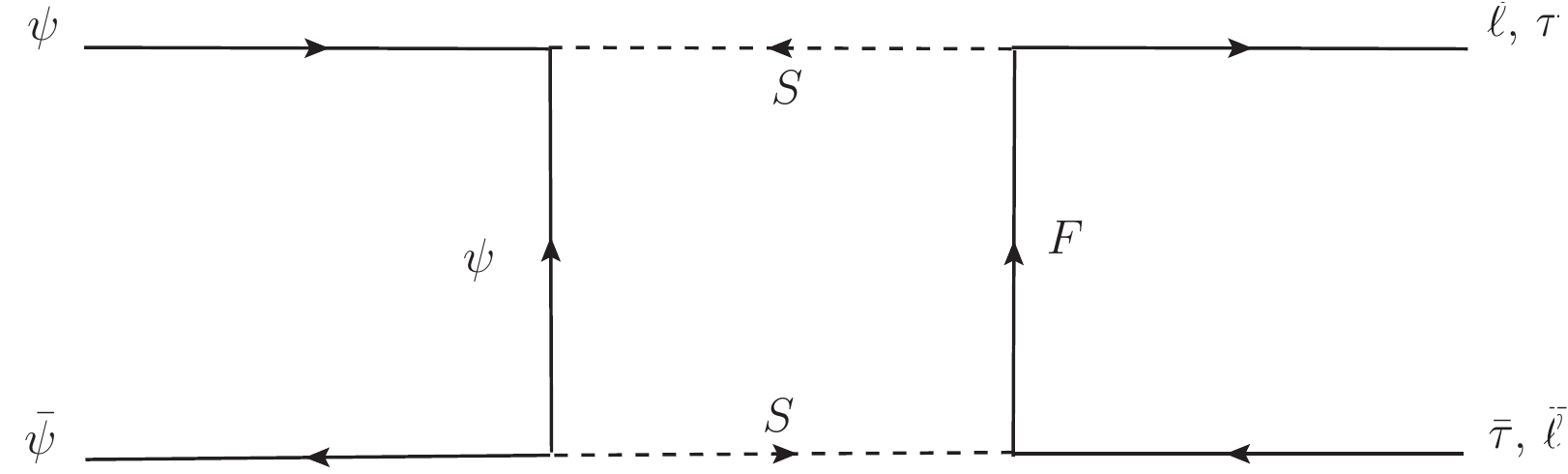}
\caption{The pNGB decay diagram. The states in the loop appear in our UV completion \eq{LagUV}.\label{fig:pi-decay}}
\end{figure}

In our UV completion, $S$ and $F$ can mediate decay of the $\Pi$ at one loop, as shown in \fig{fig:pi-decay}.
As in leptonic decays of the QCD pions, the amplitude will be proportional to the pion decay constant $F_\Pi$, which gives the strongly-coupled matrix element for $\Pi$ to vacuum.
Similarly to the $\Pi \rightarrow \bar{\nu} \nu$ decay, the spin-0 initial state requires a helicity flip for one of the final-state leptons, so that decays with at least one $\tau$ lepton are dominant.
We estimate that the overall rate goes as
\beq
\Gamma(\Pi \rightarrow \tau \ell) \sim \frac{20}{16\pi} \left( \frac{g^2 \lambda'^2}{16\pi^2} \right)^2 \left( \frac{M_\Pi^2}{M_F^2} \right)^2 \left( \frac{m_\tau}{M_\Pi} \right)^2 \frac{F_\Pi^2}{M_\Pi}. \label{eq:pi-decay}
\eeq
At our benchmark point, this yields a total decay width of $\sim 10^{-18}$ GeV, where the factor 20 takes into account the multiplicity of the amplitude and the five final states.
This is a long lifetime on detector timescales, corresponding to a decay length of $\mathcal{O}(100)$ meters.

Recall that for the dark matter relic density, we assumed that the decay width for $\Pi$ into SM final states was greater than the Hubble scale $H_\star$ at which the annihilation of $\kappa$ into other mesons freezes out.
Our benchmark values give $H_\star\sim 10^{-19}$~GeV, and $\Gamma_\Pi \sim 10^{-18}$~GeV, satisfying this condition.

In taking $\kappa$ alone to be the dark matter, we have also assumed that the heavier $\ztwo$-odd pNGB $\kappa'$ will decay promptly enough to avoid constraints from Big Bang nucleosynthesis (BBN) and other late-universe cosmology.
The radiative decay $\kappa' \rightarrow \kappa l l$, where $l$ can be a charged lepton or a neutrino, proceeds through the same operator that gives the other pion decays, but with no helicity suppression due to the three-body final state.
This process is similar to semileptonic meson decays in the SM~\cite{Ecker:1987qi}, from which we estimate that the result should be:
\beq
\Gamma(\kappa' \rightarrow \kappa l l) \sim 72 \frac{\Phi_3}{10} \left(\frac{g^2 \lambda'^2}{16\pi^2}\right)^2 \frac{M_{\kappa'}^5}{M_F^4} \left( 1 - \frac{M_{\kappa}^2}{M_{\kappa'}^2} \right)^5
\eeq
where $\Phi_3$ is a numerical factor that arises from integration over three-body phase space; we estimate it as being roughly $10^{-2}$ smaller than the two-body phase space factor of $1/16\pi$.
The numerical factor of 72 accounts for the multiplicity of final states (18) and the number of distinct amplitudes contributing (each of $\psi_1$ and $\psi_2$ appearing within the loop amplitude as in Fig.~\ref{fig:pi-decay}.)
The factor $1/10$ has been included to account for additional phase space suppression, as implied by the formalism in Ref.~\cite{Ecker:1987qi}.

At our benchmark point, we have
\beq
\Gamma(\kappa' \rightarrow \kappa l l) \sim \left(1 - \frac{M_{\kappa}^2}{M_{\kappa'}^2} \right)^{5} 10^{-20}\ \textrm{GeV}.
\eeq
For a mass difference on the order of $30\%$ between $\kappa'$ and $\kappa$, i.e. a splitting of 3 GeV, this gives a decay width of $\sim 10^{-22}$ GeV, well above the value of $\sim 10^{-24}$ GeV which would lead to active decay during BBN.

\section{Experimental signatures \label{sec:exp-signals}}

\subsection{Direct detection \label{sec:direct-detection}}
The Higgs-portal interaction in \eq{eq:pipi} yields a Higgs-exchange spin-independent direct detection cross section given by~\cite{Duerr:2015aka}
\beq
\sigma_{SI} = \frac{\xi^2\, f_n^2 \,\mu_n^2 m_n^2}{4\pi M_H^4 M_\kappa^2}
\eeq
where $\mu_n$ is the nucleon-$\kappa$ system reduced mass, $m_n$ is the QCD nucleon mass, and $f_n \sim 0.3$~\cite{Cline:2013gha} is the sigma term giving the nucleon-Higgs coupling.
Assuming that $M_\kappa \gg m_n$ so that $\mu_n \sim m_n$, we have in our model
\beq
\sigma_{SI} \sim 2 \times 10^{-12} \left(\frac{\mu_D}{\Lambda_H}\right)^2 \textrm{GeV}^{-2} \sim 10^{-39} \left(\frac{\mu_D}{\Lambda_H}\right)^2\ \textrm{cm}^2.
\eeq
This provides a significant constraint on the ratio $\mu_D / \Lambda_H$.
For our benchmark point, we have $\mu_D \sim 1$~TeV and $\Lambda_H \sim 1000$~TeV, giving $\sigma_{SI} \sim 10^{-45}$ cm${}^2$, which places our model near the current direct-detection experimental bounds from XENON1T~\cite{Aprile:2017iyp} and PandaX-II~\cite{Cui:2017nnn} for $M_\kappa \sim 10$~GeV.
This benchmark value is therefore a very interesting possibility to consider at this time and it is projected to be well within the reach of next generation direct detection experiments~\cite{Mount:2017qzi}.

\subsection{Gravitational waves \label{sec:gravity-waves}}

In the early universe, the $\suD$ sector will undergo a confining phase transition around a temperature $T_c \lsim \mu_D$.
So long as $T_c$ is larger than the electroweak transition temperature $T_{EW}\sim 100$~GeV, the confining transition will occur with all three dark fermions exactly massless.
In this limit, the $\suD$ confining transition is expected to be a first order phase transition~\cite{Pisarski:1983ms}.
Such a first-order transition is expected to produce a primordial gravitational wave signal~\cite{Schwaller:2015tja,Grojean:2006bp};
for $\mu_D$ on the order of 1-10 TeV, the frequency range of such a signal should be within the range of the proposed LISA gravitational wave observatory~\cite{Caprini:2015zlo}.

\subsection{Indirect detection \label{sec:ind-detection}}

In the present universe, the annihilation process $\kappa^\dagger \kappa \rightarrow \Pi^\dagger \Pi$ important for setting the thermal relic density will still occur, although at a rate suppressed by up to $v/c \sim 10^{-3}$ due to the scattering occurring near kinematic threshold~\cite{Buckley:2012ky} in the limit that the $\Pi$ and $\kappa$ are degenerate.
At our benchmark point, the $\Pi$ will then decay primarily into leptons as discussed above, with a preference for $\tau$ leptons.
There may be prospects for observation of a signal in gamma-ray telescopes and even precision observations of the cosmic microwave background~\cite{Madhavacheril:2013cna,Ade:2015xua}, depending on the exact model parameters used.
In particular, our benchmark scenario is close to the current exclusion limits obtained by the Planck collaboration (see Fig.~41 of Ref.~\cite{Ade:2015xua}), and should be well-probed by future observation.

\subsection{Particle colliders \label{sec:colliders}}

At the benchmark values adopted above, the Higgs can decay into the dark mesons.
For a single complex scalar, we estimate a contribution of $\Delta \Gamma / \Gamma \sim 5\%$ to the Higgs decay width within our benchmark.
Hence, we estimate
\beq
\text{Br}(H\to \Pi\,\Pi) \sim \text{Br}(H\to \kappa^{\dagger}\kappa, \kappa^{\prime \dagger}\kappa^{\prime} ) \sim 10\% \label{HiggsPiBr} \, ,
\eeq
which is well within the current upper limits on invisible Higgs decays~\cite{Olive:2016xmw,Aaboud:2017bja,Sirunyan:2017qfc}.

As noted in the dark matter discussion above, the decay width of the unstable pNGB \eq{eq:pi-decay} is fairly small, giving displaced vertices.
Moreover, the dominant decay mode of each pNGB is $\Pi \rightarrow \tau \ell$, so the final state $H \rightarrow \tau^+ \tau^+ (e/\mu)^- (e/\mu)^-$ provides a search channel with extremely low background.

It is not difficult to imagine a slight adjustment of our parameters which would yield decay lengths on the order of tens of meters or longer.
In this case, standard LHC searches would constrain $\text{Br}(H \to \Pi\,\Pi)$ only as part of the apparent branching of $H$ to invisible final states.
Exotica searches for long displaced tracks involving leptons may be sensitive up to few-meter displacements.
Direct searches for particles escaping the LHC detectors and decaying on much longer length scales, such as the proposed MATHUSLA experiment~\cite{Chou:2016lxi}, could further increase sensitivity to this exotic decay mode in the long-lifetime regime.

The UV completion also yields a large number of other composite states, including bound states of $F$ and $F'$ which are electroweak charged and may therefore be produced directly at the LHC, existing at scales of a few TeV in our benchmark model.
We leave a detailed exploration of production and decay modes of heavy composite states to future work.

Finally, as a somewhat more subtle signature, we note that confinement of $\suD$ will produce composite leptons $\Sigma^a \sim \bar{S}^a F $, which can then mix with the SM leptons through the operator $\lambda'_a S^a \bar{F} L^a \rightarrow \lambda_a' \mu_D \bar{\Sigma}^a L^a$.
Since the $\Sigma^a$ and $L^a$ have identical electroweak couplings, this mixing does not change the light lepton couplings, but it will modify the observed lepton masses slightly compared to the values generated from the Higgs mechanism.
In particular, if $m_\ell'$ is the light lepton mass generated from $y_\ell H \bar{L} \ell_R$, then we find the light mass eigenstate will be
\beq
m_\ell \sim m_\ell' \frac{M_\Sigma}{\sqrt{\lambda'^2 \mu_D^2 + M_\Sigma^2}}
\eeq
where $M_\Sigma$ is the Dirac mass of $\Sigma$.
At our benchmark point we expect $M_\Sigma \sim 2-3$ TeV, giving a modification of the lepton mass of roughly 0.1\%, or in other words, the lepton Higgs Yukawa couplings are increased by about 0.1\%.
This effect is well within current and projected experimental constraints for the lepton Yukawas in our benchmark model, but may be accessible in future high-precision measurements of $y_\tau$ and $y_\mu$ in variations of this model.

\section{Conclusions \label{sec:conclusions}}

We have presented a model in which right-handed neutrinos and dark matter are both generated as composite states belonging to the same strongly-coupled dark sector.
Introduction of an operator that violates the $U(1)$ dark baryon number symmetry naturally leads to the existence of ``dark neutrons'' with both Dirac and Majorana masses (through the analogue of neutron-anti-neutron oscillations,) suitable to give rise to neutrino masses.

The coupling to the Standard Model induced by the neutrino mass operators is too weak to lead to direct observable consequences, but dark matter direct detection can be sensitive through the Higgs portal.  The decay of the Higgs boson into dark mesons and the possibility of primordial gravitational waves give additional signatures of the dark composite sector.  In particular, these last two signatures are closely related: the Higgs can only decay into the dark mesons because they are very light due to the massless dark quarks (before electroweak symmetry breaking.)  The lightness of the dark quarks in turn is crucial to the confinement transition in the early Universe being first order.

Introduction of a simple ultraviolet completion demonstrated that the required effective operators can be generated in a straightforward way.
Moreover, the couplings within the UV completion were able to give stronger Standard Model interactions to the dark sector, leading to additional experimental signatures, including the distinctive Higgs decay $H \rightarrow \tau \tau \ell \ell'$ and the decay of heavy, charged composite states to multiple leptons.
%

\section*{Acknowledgements}
We thank Raju Venugopalan, Stefan Meinel and Taku Izubuchi for conversations about this work.
This work is supported by the U.S. Department of Energy under Grant Contracts DE-SC0012704 (H.~D. and P-P.~G.) and DE-SC0010005 (E.~N.).
E.~R. is supported by a RIKEN Special Postdoctoral fellowship.

\bibliography{arxiv_v2}

\begin{thebibliography}{26}
\expandafter\ifx\csname natexlab\endcsname\relax\def\natexlab#1{#1}\fi
\expandafter\ifx\csname bibnamefont\endcsname\relax
  \def\bibnamefont#1{#1}\fi
\expandafter\ifx\csname bibfnamefont\endcsname\relax
  \def\bibfnamefont#1{#1}\fi
\expandafter\ifx\csname citenamefont\endcsname\relax
  \def\citenamefont#1{#1}\fi
\expandafter\ifx\csname url\endcsname\relax
  \def\url#1{\texttt{#1}}\fi
\expandafter\ifx\csname urlprefix\endcsname\relax\def\urlprefix{URL }\fi
\providecommand{\bibinfo}[2]{#2}
\providecommand{\eprint}[2][]{\url{#2}}

\bibitem[{\citenamefont{Patrignani et~al.}(2016)}]{Olive:2016xmw}
\bibinfo{author}{\bibfnamefont{C.}~\bibnamefont{Patrignani}}
  \bibnamefont{et~al.} (\bibinfo{collaboration}{Particle Data Group}),
  \bibinfo{journal}{Chin. Phys.} \textbf{\bibinfo{volume}{C40}},
  \bibinfo{pages}{100001} (\bibinfo{year}{2016}).

\bibitem[{\citenamefont{Berryman et~al.}(2017)\citenamefont{Berryman,
  de~Gouv{\^e}a, Kelly, and Zhang}}]{Berryman:2017twh}
\bibinfo{author}{\bibfnamefont{J.~M.} \bibnamefont{Berryman}},
  \bibinfo{author}{\bibfnamefont{A.}~\bibnamefont{de~Gouv{\^e}a}},
  \bibinfo{author}{\bibfnamefont{K.~J.} \bibnamefont{Kelly}}, \bibnamefont{and}
  \bibinfo{author}{\bibfnamefont{Y.}~\bibnamefont{Zhang}}
  (\bibinfo{year}{2017}), \eprint{1706.02722}.

\bibitem[{\citenamefont{Agashe et~al.}(2016)\citenamefont{Agashe, Hong, and
  Vecchi}}]{Agashe:2015izu}
\bibinfo{author}{\bibfnamefont{K.}~\bibnamefont{Agashe}},
  \bibinfo{author}{\bibfnamefont{S.}~\bibnamefont{Hong}}, \bibnamefont{and}
  \bibinfo{author}{\bibfnamefont{L.}~\bibnamefont{Vecchi}},
  \bibinfo{journal}{Phys. Rev.} \textbf{\bibinfo{volume}{D94}},
  \bibinfo{pages}{013001} (\bibinfo{year}{2016}), \eprint{1512.06742}.

\bibitem[{\citenamefont{Kribs and Neil}(2016)}]{Kribs:2016cew}
\bibinfo{author}{\bibfnamefont{G.~D.} \bibnamefont{Kribs}} \bibnamefont{and}
  \bibinfo{author}{\bibfnamefont{E.~T.} \bibnamefont{Neil}},
  \bibinfo{journal}{Int. J. Mod. Phys.} \textbf{\bibinfo{volume}{A31}},
  \bibinfo{pages}{1643004} (\bibinfo{year}{2016}), \eprint{1604.04627}.

\bibitem[{\citenamefont{Gell-Mann et~al.}(1968)\citenamefont{Gell-Mann, Oakes,
  and Renner}}]{GellMann:1968rz}
\bibinfo{author}{\bibfnamefont{M.}~\bibnamefont{Gell-Mann}},
  \bibinfo{author}{\bibfnamefont{R.~J.} \bibnamefont{Oakes}}, \bibnamefont{and}
  \bibinfo{author}{\bibfnamefont{B.}~\bibnamefont{Renner}},
  \bibinfo{journal}{Phys. Rev.} \textbf{\bibinfo{volume}{175}},
  \bibinfo{pages}{2195} (\bibinfo{year}{1968}).

\bibitem[{\citenamefont{Vafa and Witten}(1984)}]{Vafa:1983tf}
\bibinfo{author}{\bibfnamefont{C.}~\bibnamefont{Vafa}} \bibnamefont{and}
  \bibinfo{author}{\bibfnamefont{E.}~\bibnamefont{Witten}},
  \bibinfo{journal}{Nucl. Phys.} \textbf{\bibinfo{volume}{B234}},
  \bibinfo{pages}{173} (\bibinfo{year}{1984}).

\bibitem[{\citenamefont{Gregory et~al.}(2012)\citenamefont{Gregory, Irving,
  Lucini, McNeile, Rago, Richards, and Rinaldi}}]{Gregory:2012hu}
\bibinfo{author}{\bibfnamefont{E.}~\bibnamefont{Gregory}},
  \bibinfo{author}{\bibfnamefont{A.}~\bibnamefont{Irving}},
  \bibinfo{author}{\bibfnamefont{B.}~\bibnamefont{Lucini}},
  \bibinfo{author}{\bibfnamefont{C.}~\bibnamefont{McNeile}},
  \bibinfo{author}{\bibfnamefont{A.}~\bibnamefont{Rago}},
  \bibinfo{author}{\bibfnamefont{C.}~\bibnamefont{Richards}}, \bibnamefont{and}
  \bibinfo{author}{\bibfnamefont{E.}~\bibnamefont{Rinaldi}},
  \bibinfo{journal}{JHEP} \textbf{\bibinfo{volume}{10}}, \bibinfo{pages}{170}
  (\bibinfo{year}{2012}), \eprint{1208.1858}.

\bibitem[{\citenamefont{Chen et~al.}(2006)}]{Chen:2005mg}
\bibinfo{author}{\bibfnamefont{Y.}~\bibnamefont{Chen}} \bibnamefont{et~al.},
  \bibinfo{journal}{Phys. Rev.} \textbf{\bibinfo{volume}{D73}},
  \bibinfo{pages}{014516} (\bibinfo{year}{2006}), \eprint{hep-lat/0510074}.

\bibitem[{\citenamefont{Soni and Zhang}(2016)}]{Soni:2016gzf}
\bibinfo{author}{\bibfnamefont{A.}~\bibnamefont{Soni}} \bibnamefont{and}
  \bibinfo{author}{\bibfnamefont{Y.}~\bibnamefont{Zhang}},
  \bibinfo{journal}{Phys. Rev.} \textbf{\bibinfo{volume}{D93}},
  \bibinfo{pages}{115025} (\bibinfo{year}{2016}), \eprint{1602.00714}.

\bibitem[{\citenamefont{Boddy et~al.}(2014)\citenamefont{Boddy, Feng,
  Kaplinghat, and Tait}}]{Boddy:2014yra}
\bibinfo{author}{\bibfnamefont{K.~K.} \bibnamefont{Boddy}},
  \bibinfo{author}{\bibfnamefont{J.~L.} \bibnamefont{Feng}},
  \bibinfo{author}{\bibfnamefont{M.}~\bibnamefont{Kaplinghat}},
  \bibnamefont{and} \bibinfo{author}{\bibfnamefont{T.~M.~P.}
  \bibnamefont{Tait}}, \bibinfo{journal}{Phys. Rev.}
  \textbf{\bibinfo{volume}{D89}}, \bibinfo{pages}{115017}
  (\bibinfo{year}{2014}), \eprint{1402.3629}.

\bibitem[{\citenamefont{Buckley and Neil}(2013)}]{Buckley:2012ky}
\bibinfo{author}{\bibfnamefont{M.~R.} \bibnamefont{Buckley}} \bibnamefont{and}
  \bibinfo{author}{\bibfnamefont{E.~T.} \bibnamefont{Neil}},
  \bibinfo{journal}{Phys. Rev.} \textbf{\bibinfo{volume}{D87}},
  \bibinfo{pages}{043510} (\bibinfo{year}{2013}), \eprint{1209.6054}.

\bibitem[{\citenamefont{Ecker et~al.}(1987)\citenamefont{Ecker, Pich, and
  de~Rafael}}]{Ecker:1987qi}
\bibinfo{author}{\bibfnamefont{G.}~\bibnamefont{Ecker}},
  \bibinfo{author}{\bibfnamefont{A.}~\bibnamefont{Pich}}, \bibnamefont{and}
  \bibinfo{author}{\bibfnamefont{E.}~\bibnamefont{de~Rafael}},
  \bibinfo{journal}{Nucl. Phys.} \textbf{\bibinfo{volume}{B291}},
  \bibinfo{pages}{692} (\bibinfo{year}{1987}).

\bibitem[{\citenamefont{Duerr et~al.}(2016)\citenamefont{Duerr,
  Fileviez~P{\'e}rez, and Smirnov}}]{Duerr:2015aka}
\bibinfo{author}{\bibfnamefont{M.}~\bibnamefont{Duerr}},
  \bibinfo{author}{\bibfnamefont{P.}~\bibnamefont{Fileviez~P{\'e}rez}},
  \bibnamefont{and} \bibinfo{author}{\bibfnamefont{J.}~\bibnamefont{Smirnov}},
  \bibinfo{journal}{JHEP} \textbf{\bibinfo{volume}{06}}, \bibinfo{pages}{152}
  (\bibinfo{year}{2016}), \eprint{1509.04282}.

\bibitem[{\citenamefont{Cline et~al.}(2013)\citenamefont{Cline, Kainulainen,
  Scott, and Weniger}}]{Cline:2013gha}
\bibinfo{author}{\bibfnamefont{J.~M.} \bibnamefont{Cline}},
  \bibinfo{author}{\bibfnamefont{K.}~\bibnamefont{Kainulainen}},
  \bibinfo{author}{\bibfnamefont{P.}~\bibnamefont{Scott}}, \bibnamefont{and}
  \bibinfo{author}{\bibfnamefont{C.}~\bibnamefont{Weniger}},
  \bibinfo{journal}{Phys. Rev.} \textbf{\bibinfo{volume}{D88}},
  \bibinfo{pages}{055025} (\bibinfo{year}{2013}), \bibinfo{note}{[Erratum:
  Phys. Rev.D92,no.3,039906(2015)]}, \eprint{1306.4710}.

\bibitem[{\citenamefont{Aprile et~al.}(2017)}]{Aprile:2017iyp}
\bibinfo{author}{\bibfnamefont{E.}~\bibnamefont{Aprile}} \bibnamefont{et~al.}
  (\bibinfo{collaboration}{XENON}), \bibinfo{journal}{Phys. Rev. Lett.}
  \textbf{\bibinfo{volume}{119}}, \bibinfo{pages}{181301}
  (\bibinfo{year}{2017}), \eprint{1705.06655}.

\bibitem[{\citenamefont{Cui et~al.}(2017)}]{Cui:2017nnn}
\bibinfo{author}{\bibfnamefont{X.}~\bibnamefont{Cui}} \bibnamefont{et~al.}
  (\bibinfo{collaboration}{PandaX-II}), \bibinfo{journal}{Phys. Rev. Lett.}
  \textbf{\bibinfo{volume}{119}}, \bibinfo{pages}{181302}
  (\bibinfo{year}{2017}), \eprint{1708.06917}.

\bibitem[{\citenamefont{Mount et~al.}(2017)}]{Mount:2017qzi}
\bibinfo{author}{\bibfnamefont{B.~J.} \bibnamefont{Mount}} \bibnamefont{et~al.}
  (\bibinfo{year}{2017}), \eprint{1703.09144}.

\bibitem[{\citenamefont{Pisarski and Wilczek}(1984)}]{Pisarski:1983ms}
\bibinfo{author}{\bibfnamefont{R.~D.} \bibnamefont{Pisarski}} \bibnamefont{and}
  \bibinfo{author}{\bibfnamefont{F.}~\bibnamefont{Wilczek}},
  \bibinfo{journal}{Phys. Rev.} \textbf{\bibinfo{volume}{D29}},
  \bibinfo{pages}{338} (\bibinfo{year}{1984}).

\bibitem[{\citenamefont{Schwaller}(2015)}]{Schwaller:2015tja}
\bibinfo{author}{\bibfnamefont{P.}~\bibnamefont{Schwaller}},
  \bibinfo{journal}{Phys. Rev. Lett.} \textbf{\bibinfo{volume}{115}},
  \bibinfo{pages}{181101} (\bibinfo{year}{2015}), \eprint{1504.07263}.

\bibitem[{\citenamefont{Grojean and Servant}(2007)}]{Grojean:2006bp}
\bibinfo{author}{\bibfnamefont{C.}~\bibnamefont{Grojean}} \bibnamefont{and}
  \bibinfo{author}{\bibfnamefont{G.}~\bibnamefont{Servant}},
  \bibinfo{journal}{Phys. Rev.} \textbf{\bibinfo{volume}{D75}},
  \bibinfo{pages}{043507} (\bibinfo{year}{2007}), \eprint{hep-ph/0607107}.

\bibitem[{\citenamefont{Caprini et~al.}(2016)}]{Caprini:2015zlo}
\bibinfo{author}{\bibfnamefont{C.}~\bibnamefont{Caprini}} \bibnamefont{et~al.},
  \bibinfo{journal}{JCAP} \textbf{\bibinfo{volume}{1604}}, \bibinfo{pages}{001}
  (\bibinfo{year}{2016}), \eprint{1512.06239}.

\bibitem[{\citenamefont{Madhavacheril et~al.}(2014)\citenamefont{Madhavacheril,
  Sehgal, and Slatyer}}]{Madhavacheril:2013cna}
\bibinfo{author}{\bibfnamefont{M.~S.} \bibnamefont{Madhavacheril}},
  \bibinfo{author}{\bibfnamefont{N.}~\bibnamefont{Sehgal}}, \bibnamefont{and}
  \bibinfo{author}{\bibfnamefont{T.~R.} \bibnamefont{Slatyer}},
  \bibinfo{journal}{Phys. Rev.} \textbf{\bibinfo{volume}{D89}},
  \bibinfo{pages}{103508} (\bibinfo{year}{2014}), \eprint{1310.3815}.

\bibitem[{\citenamefont{Ade et~al.}(2016)}]{Ade:2015xua}
\bibinfo{author}{\bibfnamefont{P.~A.~R.} \bibnamefont{Ade}}
  \bibnamefont{et~al.} (\bibinfo{collaboration}{Planck}),
  \bibinfo{journal}{Astron. Astrophys.} \textbf{\bibinfo{volume}{594}},
  \bibinfo{pages}{A13} (\bibinfo{year}{2016}), \eprint{1502.01589}.

\bibitem[{\citenamefont{Aaboud et~al.}(2017)}]{Aaboud:2017bja}
\bibinfo{author}{\bibfnamefont{M.}~\bibnamefont{Aaboud}} \bibnamefont{et~al.}
  (\bibinfo{collaboration}{ATLAS}) (\bibinfo{year}{2017}), \eprint{1708.09624}.

\bibitem[{\citenamefont{Sirunyan et~al.}(2017)}]{Sirunyan:2017qfc}
\bibinfo{author}{\bibfnamefont{A.~M.} \bibnamefont{Sirunyan}}
  \bibnamefont{et~al.} (\bibinfo{collaboration}{CMS}) (\bibinfo{year}{2017}),
  \eprint{1711.00431}.

\bibitem[{\citenamefont{Chou et~al.}(2017)\citenamefont{Chou, Curtin, and
  Lubatti}}]{Chou:2016lxi}
\bibinfo{author}{\bibfnamefont{J.~P.} \bibnamefont{Chou}},
  \bibinfo{author}{\bibfnamefont{D.}~\bibnamefont{Curtin}}, \bibnamefont{and}
  \bibinfo{author}{\bibfnamefont{H.~J.} \bibnamefont{Lubatti}},
  \bibinfo{journal}{Phys. Lett.} \textbf{\bibinfo{volume}{B767}},
  \bibinfo{pages}{29} (\bibinfo{year}{2017}), \eprint{1606.06298}.

\end{thebibliography}

\end{document}